# Towards a Conversational Measure of Trust

**Mengyao Li[1], Areen Alsaid[1], Sofia I. Noejovich[1], Ernest V. Cross[2], John D. Lee[1]**

University of Wisconsin-Madison[1]; TRACLabs[2]

mengyao.li@wisc.edu, alsaid@wisc.edu, noejovich@wisc.edu, vince.cross@traclabs.com, john.d.lee@wisc.edu

**Abstract**

The increasingly collaborative decision-making process between humans and agents demands a comprehensive, continuous, and unobtrusive measure of trust in agents. The gold standard format for measuring trust, a Likert-style survey, suffers from major limitations in dynamic human-agent interactions. We proposed a new approach to evaluate trust in a *nondirective* and *relational* conversation. The term *nondirective* refers to abstract word selections in open-ended prompts, which can probe respondents to freely describe their attitudes. The term *relational* refers to interactive conversations where respondents can clarify their responses in follow-up questions. We propose a systematic process for generating nondirective trust-based prompts by using text analysis from previously validated trust scales. This nondirective and relational approach provides a complementary trust measurement, which can unobtrusively elicit rich and dynamic information on situational trust throughout a human-agent interaction.

## Introduction

As increasingly autonomous systems are introduced into existing human-centric environments, frequent exchanges of information between humans and machines will be required to address evolving and dynamic situations (Chiou and Lee 2015; Fong et. al. 2005; Parasuraman, Sheridan, and Wickens 2000; Sun. 2006). During dynamic interactions, poor calibration of capability, barriers to interpretability, and difficulty understanding Artificial Intelligence (AI) systems can cause people distrust or overtrust (Lee and See 2004). A common solution is to have the AI explain its decisions to users. An underexplored aspect of explainable AI is the interactivity of the explanation between explainer and explainee in a conversation (Hilton 1990; Miller 2019). In this paper, we propose a nondirective and relational perspective for measuring trust through a human-agent conversation using conversational indicators of trust.

## Reconsider the Gold Standard Trust Scale

A long-established method for evaluating trust is a self-reported scale following a Likert-style survey format. Trust scales usually consist of directive statements and descriptions of human-agent relationships. For example, the frequently used trust scale in automation by Jian, Bisantz, Drury (2000) has items such as "the system is suspicious." Respondents typically record their attitudes on a continuum from 1 (not at all) to 7 (extremely). Although frequently used, this approach suffers from some limitations when assessing human-agent relationships. First, since the directive survey is heavily text-based, the administration process often forces an interruption while people are interacting with the agent. Therefore, it is hard to capture the dynamics of trust calibration that might require many administrations of the survey. Second, the direct descriptive statement does not leave respondents with adequate freedom to identify, form, and explain their feelings and opinions (Gobo 2011). For example, the statement "the system is suspicious" might cause anchoring bias, where people rely on this pre-existing information (e.g., suspicious) to judge their trust in agents. Third, most popular scales have the potential to cause positive bias in automation due to the order effect and the unbalanced design of positive-negative items (Gutzwiller *et al.* 2019). Finally, trust scales depend on and reflect human-agent relationships. Since the nature of these relationships has drastically changed in response to technological developments, past scales can be outdated and inapplicable to the inquired relationship (Merritt et al. 2019). Therefore, there is a need for an alternative or complementary trust measurement.

## A Nondirective and Relational Approach to Trust Measurement

The non-directive approach, deriving from the psychotherapy of Carl Rogers (1957), is designed to encourage and motivate respondents to provide clarifying information throughout the conversation, with indirect questions from therapists. In this paper, we adopted a nondirective approach to elicit respondents to answer open-ended questions in their own words to evaluate trust in automaton in a relational conversation.

**Nondirective conversation**

The term nondirective addresses issues on a lexical level by ensuring the choice of words in prompts are abstract and non-leading. This probes respondents to describe their own attitudes and feelings on topics of trust in automation instead of using the presumed attitudes and descriptions stated in the survey (e.g. suspicious vs. attitudes). This nondirective characteristic gives respondents more cognitive space to reflect on their trust towards the agent throughout the interactive process (Clancey 1984). A strength of this approach is that it encourages respondents to *recall* their interactions using their own language. In comparison, a survey with direct statements, where respondents *recognize* their attitudes first and then select a point to reflect their feelings, will cause respondents to reflect on their interactions even if they do not have a strong opinion.

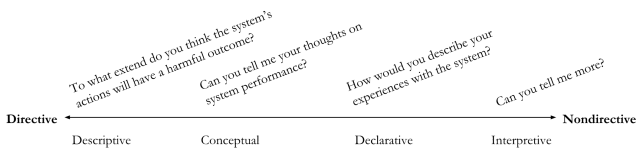

*Figure 1.* A Continuum of Levels of Prompts from Directive to Nondirective with some example questions.

A continuum of directive-nondirective prompts (i.e. descriptive, conceptual, declarative, and interpretive) is listed in Figure 1. Descriptive questions are a direct mapping of the statements in the survey (for example, "the system's actions will have harmful outcomes" turns into "to what extends do you think the system's actions will have harmful outcomes"). With descriptive questions, the specific attributes of the system are being described to respondents, which typically yields closed-ended responses. Conceptual questions involve understanding the attitudes of participants towards core concepts. For trust, these concepts would include purpose, performance, and process (Lee and See 2004). Declarative questions are used for opening questions because they usually result in an unbiased narrative from respondents. Interpretive questions are follow-up probing questions. These are used when interviewers do not fully understand respondents' answers, or the answers do not fit for the core aim of the conversational survey.

**Relational conversation**

Relational addresses issues from a conversational level. This means the responses can be clarified in an interactive manner with a conversational agent, instead of being forced through a one-shot, closed-ended, point-based system (Chiou n.d.). The relational characteristic can capture more dynamic and rich information in an unobtrusive manner. In comparison to only numeric survey data, the conversation information contains content delivered by vocal cues, conversational turn-taking, sequences of attitude, and conversational ending.

A relational conversation can also provide an unobtrusive assessment of trust while the user operates the primary task of an experiment, such as driving. The continuity of the conversation promises the integration of a conversational agent into the experimental system, which allows trust to be monitored dynamically. By employing this method, we can capture situational trust throughout various scenarios including both *external* variability (i.e. complexity of a system, task difficulty, workload, perceived risks and benefits) and the internal variability of the people (i.e. self-confidence, subject matter expertise, familiarity, attentional capacity) (Hoff, Bashir 2015). The relational approach provides a means for building bidirectional trust e.g., provides states, goals, attitudes, such as trust indicators, from human to agent, and gains insights about system capability from agent to human (Ezer et. al. 2019).

To structure the relational conversation, we propose to start with declarative questions and follow with conceptual questions with the slot-filling technique to probe theoretical concepts that must be evaluated. Throughout the survey, the interviewee may ask interpretative and/or descriptive questions if needed.

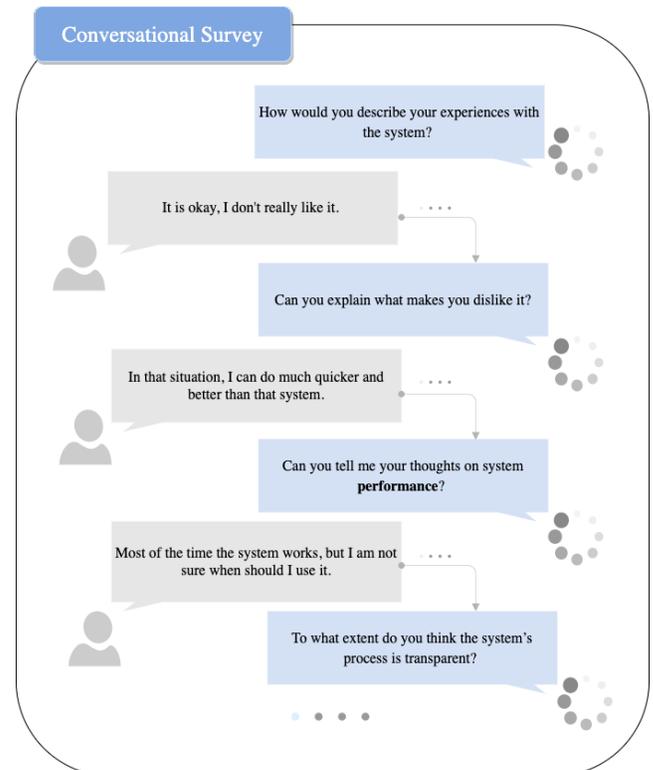

*Figure 2.* An example of conversational survey.

An example of the proposed conversational survey with nondirective prompts is shown in Figure 2. The conversation starts with declarative question. Positive and negative intents are designed and classified in the agent conversational architecture. If the negative intent is detected, shown as the user's first response in the figure ("I don't really like it"), the follow-up interpretative question would occur ("Can you explain what makes you dislike it?"). Once the user answers the follow-up question, the agent would initiate the conceptual question in the formulated prompts based on slot-filling ("Can you tell me your thoughts on system performance?"). Another descriptive question is followed up. The nondirective characteristic reflects in the lexical selection highlighted in the figure (e.g. system performance). The relational characteristic reflects the back and forth conversation with the designed follow-up intents. This example is a simple running demo to show a novel perspective of relational and nondirective approach. Some aspects of this approach that are still in development include the formulation of additional scenarios, the architecture of the agent, the process for response decoding, and the analysis method. The planned studies to address these aspects are discussed in the Future Work section.

In this paper, we discussed some limitations of the traditional trust scale and proposed a relational and nondirective approach as a supplementary method to measure trust called, *conversational survey*, which can measure situational trust dynamically and unobtrusively. Similar, to a structured interview, we incorporate a conversational chatbot and text analysis techniques to develop a relational and nondirective approach to measure trust. The methodology of conversational survey of trust requires prompt generation, response decoding, analysis, and scoring. For this paper, due to space limitations and project progress, we only outlined the first step in the method: prompt generation. The relational and nondirective prompts can provide a novel and meaningful perspective to elicit and measure trust, especially for dynamic relationships between humans and agents.

## Elements of a Conversational Measure of Trust

### Basis for Trust Prompts

A thorough review of trust scales was conducted by Alsaid, Lee, and Chiou (2020). The authors categorized trust assessment scales into three domains (i.e., trust in automation, trust in e-commerce, and trust in humans) and three temporal compositions (i.e., dispositional trust, history-based trust, and situational trust). The analysis identified similarities between the trust scales at the level of the scale as a whole, items within scales, and words encompassed in items. Their results provide insight into the common words used to assess trust in general, and in each domain. In our proposed method, we built on their findings to construct nondirective prompts in conversational trust surveys that can be used to evaluate trust through conversations.

### Text Analysis to Identify Trust Prompts

To provide nondirective trust prompts, it is essential to analyze trust-related words and their latent meanings. Rather than taking a qualitative approach (e.g., open coding or axial coding), which is subjective, laborious, hard to replicate, and dependent on expertise in theoretical understanding of trust, we used text analysis to extract nondirective words and form trust prompts.

### Word Embedding

A key idea in the examination of textual data is representing words as numeric vectors, known as *embeddings*. Word embeddings describe words in vectors in a high-dimensional semantic space based on their co-occurrence of words within a small chunk of text across a large corpus of documents. Numerous methods can be used to train estimate word embeddings (Goldberg and Levy 2014; Pennington et. al. 2014). For more details on theoretical comparisons between methods, see Alsaid, Lee, and Chiou (2020). In our paper, we use the Global Vectors for Word Representation (GloVe) method, developed by Pennington, Socher, and Manning (2014). GloVe is an embedding technique based on factorizing a matrix of word co-occurrence statistics, which shows improved interpretability and accuracy.

### Hierarchical clustering

Hierarchical clustering is a method of cluster analysis for identifying groups in a dataset. It can result in a tree-based representation of data, called a dendrogram, which shows the hierarchical relationship between clusters. The horizontal axis of the dendrogram represents the distance or dissimilarity between clusters. The vertical axis represents the clusters. Each node represents the joining or fusion of two clusters by the splitting of a horizontal line into two horizontal lines on the graph. The horizontal position of the split, shown by the short vertical bar, gives the distance (dissimilarity) between the two clusters. Using the numeric representations of trust words from the word embedding, we can create a hierarchy of trust-related word clusters. Based on the dendrogram representation of trust words and methods for text summarization, the high-level node information can be mapped to the nondirective words to capture its lower level trust words.

## Method

To generate the nondirective prompt, we first composed the dataset using exact wordings in the surveys identified by Alsaid, Lee, and Chiou (2020), which are highly influential trust scales (see Figure A. 1 and Figure A. 2). Then we processed the data by converting all words in lower case, removing stop words and stemming the words to extract the base form. A ten-step procedure is outlined below:

1. *Identify your domain*

Differences in domains would result in the various definitions and measurement focus of trust. The trust scales can be categorized by three domains: human by accessing the interpersonal trust; E-commerce by accessing consumers' trust in brand; and automation by accessing trust in technology, automation, robot, intelligent agents generally. For the scope of this paper, we would filter scales pertaining to the automation domain based on scales selected in Alsaid, Lee, and Chiou (2020).

2. *Rank domain-specific words*
   The log odds ratio of words representing the proportion of domain-related words in each scale is shown in Alsaid, Lee, and Chiou (2020) (see Appendix Figure A. 1). Using filtered scales based on our focused domain from step 1, we can rank scale in Figure A. 1 based on domain-related percentages of trust words in automation, from high-to-low. This step can give us the scale ranking by the "most significant terms to distinguish trust" in this specific domain.

3. *Rank domain-specific citations*
   Then rank the similar scales based on the number of citations the papers have received to highlight the more popular 9 scales in the Figure A. 1. The number selected for filtering the scales can be changed based on the size of the original dataset and the words needed for prompt development. The larger the number, the more items are included in further analysis. Although this approach is more inclusive, more items will tend to increase variance.

4. *Rank by construct-specified words*
   Similar to domain related words, differences in trust construct would result in the different wording selections for trust scales. The trust can be categorized by three constructs: dispositional measuring people's general tendency to be complacent and trusting; history-based measuring people's trust based on involving interactions and experiences; situational measuring trust in a specific scenario or context. Since the conversational characteristics can dynamically capture the scenario variability, we would focus on situational trust measurement for this paper. Rank Figure A. 2 based on the selected categories of the trust construct, situational, from high-to-low.

5. *Rank construct-specific citation*
   Based on results from step 4, rank the similar scales based on citation and highlight the top 3 scales in the targeted category (i.e. situational trust).

6. *Generate database for prompt generation*
   Take a union set of results based on step 3 and step 5 and list out all items of each scale. Then, add any additional validated scales that are not included in the original dataset when it is published.

7. *Calculate the word embedding*
   Combine all items as bag of words to conduct the word embedding. Transform words into a numeric representation.

8. *Perform hierarchical clustering*
   Apply hierarchical clustering based on step 8 results to form the dendrogram showing the hierarchical relationship between trust words.

9. *Select hierarchical word representation of trust*
   Based on the dendrogram, identify the threshold to 'cut the tree', which means to define how many clusters of words to keep. We adopted the centroid-based text summarization of the word embeddings in each cluster (Rossiello, Basile, and Semeraro, 2017). Using the centroid information, we can map its word embedding into the nondirective terms eliciting trust.

10. *Formulate non-directive prompts*
    Follow the adjusted guidelines of nondirective approach and correspond to the continuum of directive-nondirective prompts in Figure 1 to form the structured questions and unstructured prompts in the conversation to elicit trusting (Josefi and Ryan 2004).

The 10-step procedure described has two functions: 1) steps 1–6 researchers provide guides the selection of scales and develop trust measurement based on existing ones; 2) Steps 7–10 are generate nondirective conversational prompts. The result from dendrogram in step 9 shows the hierarchical word presentation and summarization of trust, which can be used to extract the inclusive and nondirective words to identify lower-level trust-related words.

## Future Work

The methodology of conversational measure of trust consists of a process of prompts generation, responses decoding, analysis, and scoring. In this paper, we only presented a general method framework for the first part: nondirective prompt generation. Additionally, although conversational survey promises to give researchers rich information, it is only meaningful when analyzed appropriately. Results of prompt generation, the continuing response decoding, analysis, and scoring are still in process. Detailed processes and evaluation of text augmentation will be provided in the future study. We are also interested in further analyzing the role issues of the conversational agent: whether to administrate trust in a close-loop system through first-person narrative as a closed-loop system, or separate from the target system, or administrate trust questions through third-person narrative. Limitation for the conversational measurement approach is that it is not suitable for all conditions, such as high workload, limited time window for administration, or when auditory channel is blocked.

Using the prompts generated following the process in this paper, we will conduct a study involving resource management and negotiation with a conversational agent. The generated conversational exchanges between human and agents provide data for extracting and analyzing trust indicators. For the responses decoding and analysis, a three-level (i.e. micro, meso, macro) analysis by extracting the trust-related responses through the conversations will be conducted. A strategy called triangulation, which uses corroborating evidence from multiple perspectives, can increase confidence in the validity of the trust measurement (Lazar, Feng, Hochheister 2017). The end goal of this novel conversational survey method is to provide an alternative unobtrusive measurement or administrate along with the traditional text-based scales. We hope this new measure can unobtrusively elicit rich and dynamic information on situational trust throughout a human-agent interaction.

# Acknowledgement

We thank members of the University of Wisconsin-Madison Cognitive Systems Laboratory for their insightful discussions and comments. This research project is funded by NASA 's Human Research Program (HRP).

# References

Alsaid, A.; Lee, J.D.; Chiou, E. ca.2020., Understanding similarities and differences across trust questionnaires: A text analysis approach. To be submitted to *ACM Transactions on Computer-Human Interaction*.

Clancey, W.J. 1984. Classification Problem Solving. In *Proceedings of the Fourth National Conference on Artificial Intelligence*, 49-54. Menlo Park, Calif.: AAAI Press.

Chiou, E. K. n.d. Trusting Automated Agents: Designing for Appropriate Cooperation. Human Factors.

Chiou, E. K., and Lee, J. D. 2015. Beyond reliance and compliance: human-automation coordination and cooperation. *In Proceedings of the Human Factors and Ergonomics Society Annual Meeting* 59(1): 195 –199. Sage CA: Los Angeles, CA: SAGE Publications.

Ezer, N.; Bruni, S.; Cai, Y.; Hepenstal, S. J.; Miller, C. A.; Schmorrow, D. D. 2019. Trust Engineering for Human-AI Teams. *In Proceedings of the Human Factors and Ergonomics Society Annual Meeting* 63(1): 322–326. Sage CA: Los Angeles, CA: SAGE Publications.

Fong, T.; Nourbakhsh, I.; Kunz, C.; Fluckiger, L.; Schreiner, J.; Ambrose, R.; Burridge, R.; Simmons, R.; Hiatt, L., Schultz, A.; Trafton, J.G. 2005. The peer-to-peer human-robot interaction project. *In Space 2005:* 6750.

Gobo, G. 2011. Back to likert: towards the conversational survey. *The SAGE handbook of innovation in social research methods:* 228-248, London: SAGE Publications Ltd

Gutzwiller, R.S.; Chiou, E.K.; Craig, S.D.; Lewis, C.M.; Lematta, G.J.; Hsiung, C.P. 2019. November. Positive bias in the 'Trust in Automated Systems Survey'? An examination of the Jian et al. (2000) scale. *In Proceedings of the Human Factors and Ergonomics Society Annual Meeting* 63 (1): 217–221. Sage CA: Los Angeles, CA: SAGE Publications.

Lazar, J.; Feng, J. H.; Hochheiser, H. 2017. *Research methods in human-computer interaction.* Morgan Kaufmann.

Lee, J. D. and See, K. A. 2004. Trust in automation: Designing for appropriate reliance. *Human factors* 46(1): 50-80.

Goldberg, Y. and Levy, O. 2014. Dependency-based word embeddings. *In Proceedings of the 52nd Annual Meeting of the Association for Computational Linguistics* 2: 302–308.

Merritt, S.M.; Ako-Brew, A.; Bryant, W.J.; Staley, A.; McKenna, M.; Leone, A.; Shirase, L. 2019. Automation-induced complacency potential: Development and validation of a new scale. *Frontiers in psychology*, 10: 225.

Miller, T. 2019. Explanation in artificial intelligence: Insights from the social sciences. *Artificial Intelligence*, 267: 1–38.

Hoff, K.A. and Bashir, M. 2015. Trust in automation: Integrating empirical evidence on factors that influence trust. *Human factors* 57(3): 407–434.

Hilton, D.J.; 1990. Conversational processes and causal explanation. *Psychological Bulletin* 107(1): 65.

Jian, J. Y.; Bisantz, A. M.; Drury, C. G.; 2000. Foundations for an empirically determined scale of trust in automated systems. *International journal of cognitive ergonomics* 4(1): 53– 71.

Josefi, O. and Ryan, V., 2004. Non-directive play therapy for young children with autism: A case study. *Clinical Child Psychology and Psychiatry 9*(4): 533 –551.

Parasuraman, R.; Sheridan, T. B.; Wickens, C. D. 2000. A model for types and levels of human interaction with automation. *IEEE Transactions on systems, man, and cybernetics-Part A: Systems and Humans* 30(3): 286–297.

Pennington, J.; Socher, R.; Manning, C. D. 2014. Glove: Global vectors for word representation. *In Proceedings of the 2014 conference on empirical methods in natural language processing*.

Rogers, C. R. 1957. The necessary and sufficient conditions of therapeutic personality change. *Journal of consulting psychology 21*(2): 95.

Rossiello, G.; Basile, P.; Semeraro, G. 2017. Centroid-based text summarization through compositionality of word embeddings. *In Proceedings of the MultiLing 2017 Workshop on Summarization and Summary Evaluation Across Source Types and Genres:* 12–21.

Sun, R. 2006. *Cognition and Multi-Agent Interaction: From Cognitive Modeling to Social Simulation.* Cambridge University Press.

# Appendix

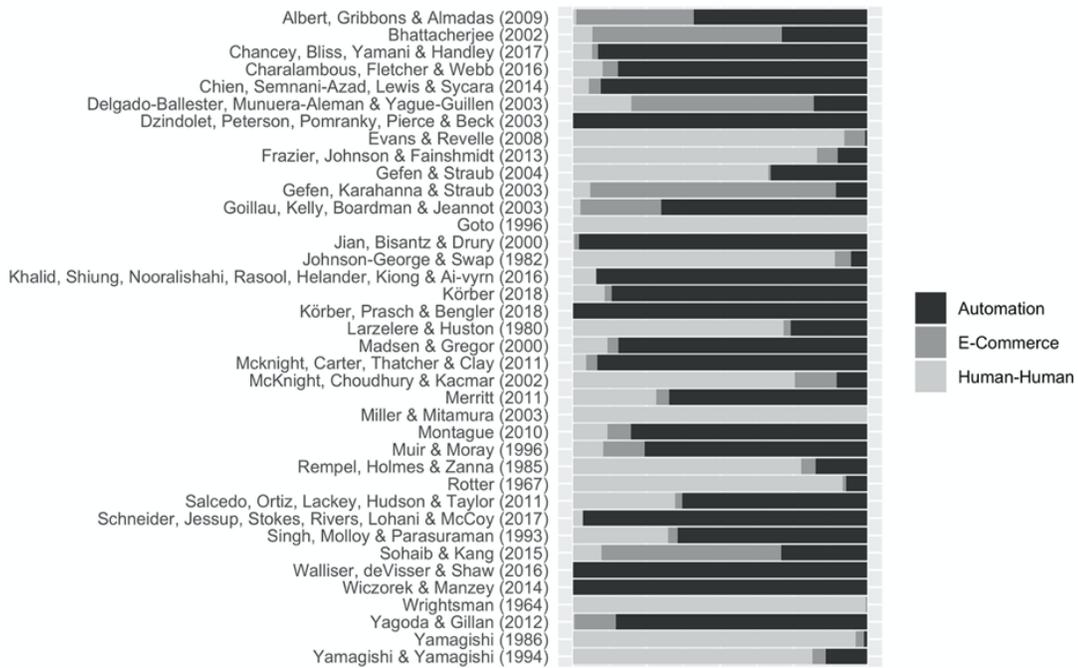

*Figure A. 1.* Trust questionnaires' domain composition (containing words most related to which domains of trust).

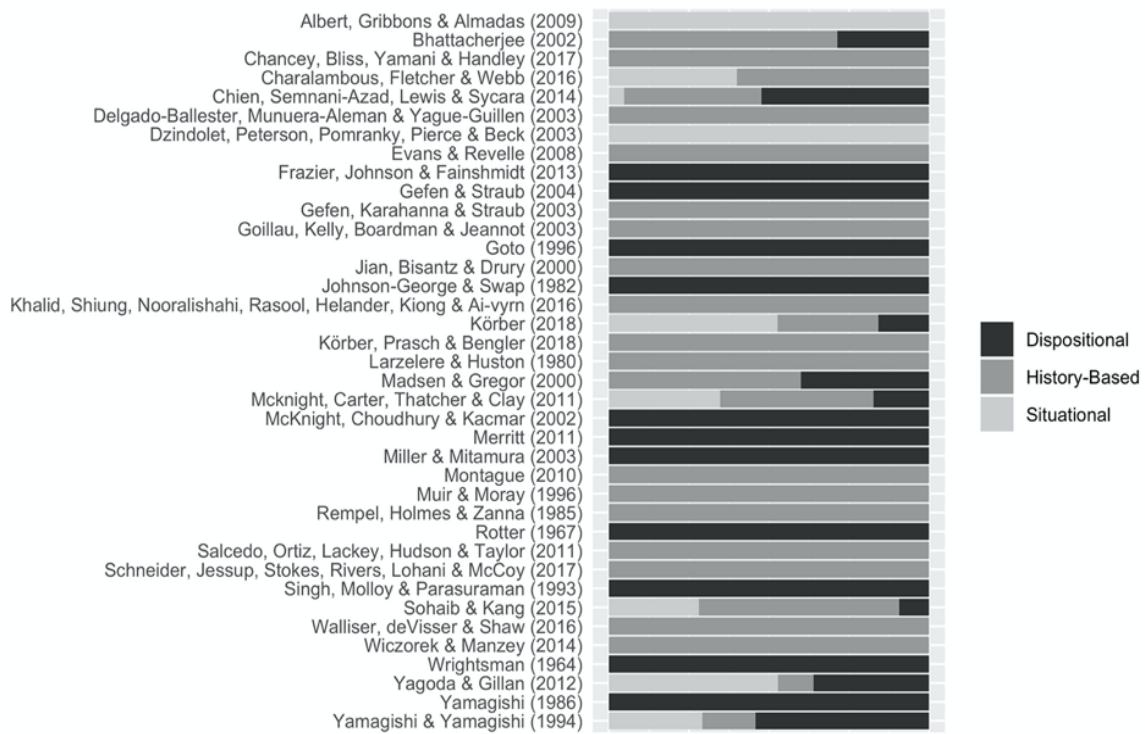

*Figure A. 2.* Trust questionnaires' temporal composition of each questionnaire item.